\documentclass[11pt]{article}
\usepackage{graphicx}
\usepackage{epsfig}
\usepackage{amssymb,amsmath}
\usepackage{graphicx}  
\newcommand{\beq}{\begin{equation}}
\newcommand{\eeq}{\end{equation}}
\newcommand{\be}{\begin{equation}}
\newcommand{\ee}{\end{equation}}
\newcommand{\bea}{\begin{eqnarray}}
\newcommand{\eea}{\end{eqnarray}}

\fontencoding{T1}
\fontfamily{garamond}
\fontseries{m}
\fontshape{it}
\fontsize{13}{15}
\selectfont

\begin{document}

\title{Counting valence quarks at RHIC and LHC}
\author{L.~Maiani$^a$, A.D. Polosa$^b$, V.~Riquer$^b$, C.A.~Salgado$^a$\footnote{
Permanent Address: Departamento de F{\'\i}sica de Part{\'\i}culas, Universidade
de Santiago
de Compostela (Spain).}\\
$^a$Dip. Fisica, Universit\`a  di Roma ``La Sapienza'' and INFN, Roma, Italy\\
$^b$INFN, Sezione di Roma, Roma, Italy}
\maketitle
\begin{abstract}
We consider the Nuclear Modification Ratios in heavy ion collisions, R$_{CP}$ and R$_{AA}$, in the region of intermediate transverse momentum, and study the dependency upon the constituent quark composition of the observed hadron. Adopting a two component recombination/fragmentation model, validated by experimental information from STAR and PHENIX, we show that a clear 
distinction is predicted for the $f_0(980)$ between the assumptions of $s\bar s$ or diquark-antidiquark content.
\newline
\newline
\newline
{\bf Preprint No.} Roma1-1432/2006 \newline
\end{abstract}
\newpage

\section{Introduction}

Some of the most important results obtained so far at RHIC are related with the high-$p_\perp$ spectrum in heavy ion collisions. A very strong suppression of high-$p_\perp$ particles \cite{Adcoxjp1} 
is commonly interpreted in terms of energy loss of the energetic partons when traversing a very opaque medium (see e.g. \cite{Salgado:2005pr,reviewart1} for  recent reviews); two particle correlations offer the first jet-like measurements in heavy ion collisions, with some surprising features which await explanation 
\cite{Adler:2005ee, Adams:2006yt}.

On the other hand, unexpectedly, the protons yield dominates over the pion yield at intermediate values of 1 GeV$\lesssim p_\perp\lesssim$ 6 GeV \cite{Adler:2006xd}, in contrast with any other result in experiments involving hadrons. This increase in the relative yield of protons is in fact common to other baryons and the medium effects in this $p_\perp$-region have been found to be driven by the number of constituent quarks of the particle studied. In particular, the amount of suppression has been experimentally found to depend on quark counting rather than on particle mass. Similarly, the azimuthal momentum anisotropies, measured by the second order, $v_2$, of a harmonic expansion in the azimuthal angle, were found to exhibit scaling with the particle quark number (for a recent review of the data see, e.g.~\cite{tserruya}). An economical model to describe these features of intermediate-$p_\perp$ particles is the recombination or coalescence model \cite{Friesvb,Greco:2003mm}. 

The overall picture is that one exploits the different behavior exhibited by parton fragmentation in the hard sector and by hadronization in the soft sector: the cost to produce a particle in the fragmentation of a high-$p_\perp$ parton increases with the number of constituent quarks, while recombination gives essentially the same spectrum due to its additive nature in momentum.  Due to the dominance of the recombination component, baryon relative yields are enhanced in central $A+A$ collisions with respect to peripheral collisions or $p+p$ collisions. Predictions for multicharmed hadron abundances in the coalescence regime have been recently presented in ~\cite{becattini}.

Other models, based on a hydrodynamical description of a heavy ion collision describe reasonably the soft part of the spectrum (see e.g. \cite{Eskola:2005ue}). Models based on baryon formation mediated by string-junction configurations can also account for the observed enhancement in the ratio baryon/meson \cite{Pop:2004dq}. These type of models have been applied before to SPS data to explain the strong enhancement of strange baryon production \cite{Capella:1999cz, Vance:1999pr}. Understanding the dynamical origin of these effects, which suggest a strong interplay between hadronization, thermalization and parton shower evolution, needs of additional observables.

The validity of the coalescence model in a sizeable range of $p_\perp$ may serve a dual purpose. It confirms the formation of an unconfined state in central heavy ion collisions and it opens up the exciting possibility to use the unconfined state to probe the constituent quark composition of hadrons, thus discriminating between conflicting models. One proposal in this direction \cite{Nonakaew} was to measure the azimuthal anisotropy of the possible resonances which were interpreted as pentaquarks a couple of years ago. The difficulty is that, on one hand these resonances -- if real resonances at all -- are difficult to measure and, on the other hand, the azimuthal anisotropies is a very small, percent level, signal. 

In the present paper, we consider the nuclear modification ratios, $R_{CP}$ and $R_{AA}$, as experimental observables with large signals, and find that they are suitable to {\it measure} the valence quark content of the resonances and allow to clarify their nature.

We take, in particular, the case of the $f_0(980)$, proposed to be a 4-quark meson in~\cite{scalars1} and already measured in heavy ion collisions \cite{Adams:2003cc}. Being a well known resonance, abundantly produced and easy to identify, the $f_0(980)$ provides a unique opportunity for these types of studies. 

We consider the two alternatives; $f_0(980)= s\bar s$ \cite{aandcat} or $f_0(980)= [\bar q \bar s][qs]$ \cite{scalars1} and show that they can be very well distinguished by data at RHIC. We consider the extension of our analysis at LHC and find rather similar results.
Other interesting cases are the $f_0(1370)$ and $f_0(1710)$ for which interpretations as $q\bar q$ or tetraquark states have been advanced, in \cite{frankclose1} and in \cite{noilast} respectively. 

According to the tetraquark interpretation, $f_0(1710)$ has the same composition as $f_0(980)$. The  $R_{CP}$ and $R_{AA}$ predictions for this state are in fact essentially the same as for $f_0(980)$.  
The $f_0(1370)$ is supposed to be a state $[nn][\bar n\bar n]$ ($n=u,d$),  a recurrence of the $\sigma(500)$ meson in the multiquark scheme. A brief discussion of the expected signal for $f_0(1370)$ is given at the end of this paper.
An open problem in this context is how a glueball hadron would scale, e.g the $f_0(1500)$ as proposed in \cite{frankclose1,noilast}. Also lacking are predictions for a molecular state, such as the $f_0(980)$ in \cite{molecule}.

\section{The model}
\label{model}

The soft part of the spectrum is assumed to be formed by the decay of a deconfined state of constituent quarks which recombine to form the observed hadrons, following valence quark counting. The hard part is described by perturbative QCD with a simple implementation of the energy loss of the high energy partons traversing the medium. 

We summarize in this Section basic formulae and parameters used for the two components of the hadron production model, recombination and fragmentation, following Ref. \cite{Friesvb}. Notice that gluon degrees of freedom are not included in the original model, they have been discussed in \cite{Muller:2005pv} but the results are basically unchanged and will not be considered here. 

\paragraph{Recombination.} 

The basic ingredients are the constituent quark or antiquark  distribution functions, which we write according to:
\beq
\label{quarkdist}
E_a\frac{d\delta N_a}{d^3p_a}=\delta\Omega \cdot \gamma_a \cdot \rho_a(p_a, T)\cdot f_a(p_a, \phi_a)\;,
\eeq
where $\gamma_a$ is the fugacity of the constituent and $f_a$ is a function that takes into account the azimuthal dependence of the distribution for non-central collisions. For recombination from a non-relativistic fluid at temperature $T$, $\delta\Omega$ would simply be  $E_a dV$, $dV$ being
the volume element, and $\rho_a$ the Boltzmann distribution. For the case we consider, namely recombination from a relativistic, boost-invariant fluid \cite{bj83}, $\delta\Omega$ is proportional to the volume element of the 3-dimensional, boost-invariant hypersurface defined in 4-space by:
\bea
&&x^\mu=(t,~x,~y,~z)\nonumber \\
&&\tau^2=t^2-z^2={\rm const.};~t=\tau \cosh\eta,~z=\tau \sinh\eta  \nonumber \\
&&u^\mu={\rm normal~to~hypersurface}=(\cosh\eta,0_\perp,\sinh\eta)\;.
\label{3dimspace}
\eea
Accordingly:
\beq
\delta \Omega=\tau d\eta~d^2x_\perp (u\cdot p_a)\;.
\eeq
and $x_\perp=(x,~y)$ is the 2-dimensional transverse space. 

The Boltzmann distribution is replaced by the Touschek \cite{touschek} invariant phase-space distribution:
\beq
e^{-\frac{E_a}{T}}~\to e^{-\frac{(v\cdot p_a)}{T}}\;,
\eeq
We denote by $v^\mu$ the local 4-velocity of the fluid and by $v_\perp$ the transverse expansion velocity:
\bea 
&& v^\mu=(\cosh \eta \cosh \eta_\perp,~\sinh \eta_\perp \cos\phi,~\sinh \eta_\perp \sin\phi,~\sinh \eta \cosh \eta_\perp)\nonumber \\
&&v_\perp=\tanh\eta_\perp\;.
\label{radialflowvel}
\eea

Denoting by $y_a$ and $\phi_a$ the constituent rapidity and azimuth in the transverse plane and by $m_{a,\perp}=\sqrt{m_a^2+p_\perp^2}$ the transverse mass, we have:
\beq
(v\cdot p_a)=m_{a,\perp} \cosh\eta_\perp \cosh (\eta-y)-p_\perp \sinh\eta_\perp \cos(\phi-\phi_a)\;.
\eeq 

In the coalescence component, hadrons with transverse momentum $P_\perp$ are formed by recombination from the thermal spectrum of partons, which share the momentum of the hadron equally, in the most favorable situation. 

We specialize, for simplicity, to the case of a $q\bar q$ meson and use the notations of Ref. \cite{Friesvb}, hadron variables are denoted with capital letters. The  meson momentum distribution at mid-rapidity is given by:
\bea
&&\left(E_M\frac{dN_M}{d^3P}\right)_{y=0}=\left(\frac{dN_M}{dyd^2P_\perp}\right)_{y=0}=\nonumber \\
&&=\gamma_a\gamma_bC_M\int \tau d\eta~r_\perp dr_\perp d\phi ~(u\cdot P)~\left[\rho_a(\frac{P_\perp}{2},T)\rho_b(\frac{P_\perp}{2},T)\right]~\left[ f_a(\frac{P_\perp}{2}, \Phi)f_b(\frac{P_\perp}{2}, \Phi)\right]\;,
\label{hdistr}
\eea
where $C_M$ is a multiplicity factor and $P^\mu$ the hadron momentum. 

We first neglect the azimuthal dependence by setting $f_a=f_b=1$. 
The integrals over $\eta$ and $\phi$ give rise to approximate exponentials and one finds, in conclusion \cite{Friesvb}:
\begin{equation}
  \frac{dN_M}{d^2 P_\perp dy}\Big|_{y=0} = 
  C_M M_\perp \frac{\tau A_\perp}{(2\pi)^3} \, 2\gamma_a \gamma_b \,
  I_0 \left[ \frac{P_\perp \sinh \eta_\perp}{T}\right] 
  k_2(P_\perp) \, ,
  \label{eq:messpec}
\end{equation}
$A_\perp=\pi \rho_0^2$ is the transverse area of the partonic medium at freeze-out, $\tau$ the hadronization time and $M_\perp$ is the transverse mass of the hadron. We have introduced the Bessel functions:
\begin{eqnarray}
&&I_0(z)=\frac{1}{\pi}\int_0^\pi d\phi ~e^{-z\cos\phi};~~K_1(z)=\int_0^\infty d\eta \cosh\eta~ e^{-z\cosh\eta}
\end{eqnarray}
and the short-hand notation for a N-quark hadron:
\begin{eqnarray}
  k_N(P_\perp) &=&  
  K_1 \left[ \frac{\cosh \eta_\perp}{T}\sum_{a=1}^N\sqrt{m_a^2 + \frac{P_\perp^2}{N^2}}\right] \, .
\label{eq:k4q}
\end{eqnarray}

For baryons and 4-quark states we find, similarly:
\begin{equation}
  \frac{dN_B}{d^2 P_\perp dy}\Big|_{y=0} = 
  C_B M_\perp \frac{\tau A_\perp}{(2\pi)^3} \, 2\gamma_a \gamma_b \gamma_c \,
  I_0 \left[ \frac{P_\perp \sinh \eta_\perp}{T}\right] k_3(P_\perp)
  \label{eq:barspec}
\end{equation}
\begin{equation}
  \frac{dN_{4q}}{d^2 P_\perp dy}\Big|_{y=0} = 
  C_{4q} M_\perp \frac{\tau A_\perp}{(2\pi)^3} \, 2\gamma_a \gamma_b \gamma_c \gamma_d \,
  I_0 \left[ \frac{P_\perp \sinh \eta_\perp}{T}\right] k_4(P_\perp)\;.
  \label{eq:4qspec}
\end{equation}

Following \cite{Friesvb} the degeneracy factors are taken to account for the spin of the final hadron only and we do not distinguish between $\Lambda$ and $\Sigma^0$, so that: $C_M=1$, $C_B=2$ for protons, $C_B=4$ for $\Lambda +\Sigma^0$ and $C_{4q}=1$.

Until now we have considered central collisions. For peripheral collisions a further, common fugacity factor for all parton species $\gamma_{{\rm periph}}$, is included and the transverse size of the medium is scaled with the impact parameter, $b$, according to: 
\beq
\rho_0^2 \to \rho_0^2~\frac{ \sqrt{R_A^2-\frac{b^2}{4}}~(R_A-\frac{b}{2})}{R_A^2}
\eeq
where $R_A$ is the radius of the nucleus. 

We list in Table~\ref{table1} the values of the parameters assumed for central and peripheral collisions at RHIC. We follow here mostly Ref.~\cite{Friesvb}, where it is shown that these parameters reproduce reasonably the intermediate $p_\perp$ cross sections at RHIC.
\begin{table}[htb]
\caption{{\footnotesize Parameters of the hadron production model in Au+Au @ RHIC, Pb+Pb @LHC and p+p. Numbers in parentheses in the last column refer to peripheral collisions with $b=12$~fm at RHIC and $b=14$~fm at LHC.}}
\begin{center}
\label{table1}
\begin{tabular}{@{}|c|c|c|c|c|c|c|c|c|}
\hline
~~~~               & $\gamma_{u,d}$ & $\gamma_{\bar u,\bar d}$ & $\gamma_{s,\bar s}$ &$\gamma_{{\rm periph}}$ & $\begin{array}{c} \tau A_T \\ ({\small {\rm fm} ^3)}\end{array}$ &v$_\perp$& $\begin{array}{c}\epsilon_0\\({\rm GeV}^{-1/2})\end{array}$ &N$_{coll}$ \\
\hline
Au+Au @RHIC       	& 1	& 0.9 & 0.8 & 0.7 &  1.27$\cdot 10^3$ & 0.55& 0.82&1146 (26)  \\
\hline
Pb+Pb @LHC          & 1	& 1 & 1 & 0.7 & 11.5$\cdot 10^3$ & 0.68 & 2.5 & 3643(82) \\
\hline
p+p @RHIC(LHC)		&0.4 &  0.4 & 0.12 (0.4)& --  & 13.2 (119) & 0.55 (0.68) & 0& -- \\
\hline
\end{tabular}\\[2pt]
\end{center}
\end{table}

To extend the recombination mechanism to the LHC, we need criteria to extrapolate the parameters of the model. In absence of data, we compare with the theoretical results obtained with the hydrodynamical approach in \cite{Eskola:2005ue}. We find that,  by  rescaling $v_\perp$ and $\tau A_\perp$ to the values reported in the second line of Table~\ref{table1}, a reasonable agreement is obtained for slope and normalization of the predicted yields for $\pi$, $p$ and $K$. We also take unit fugacities for central collisions, although this has a negligible effect in the final result. In \cite{Fries:2003fr} different values for the extrapolation of the value of $v_\perp$ from RHIC to LHC have been found to affect the
final result for the suppression. Here we take the conservative approach of fitting hydrodynamical results.

\begin{figure}
\begin{minipage}{0.32\textwidth}
\begin{center}
\includegraphics[width=\textwidth]{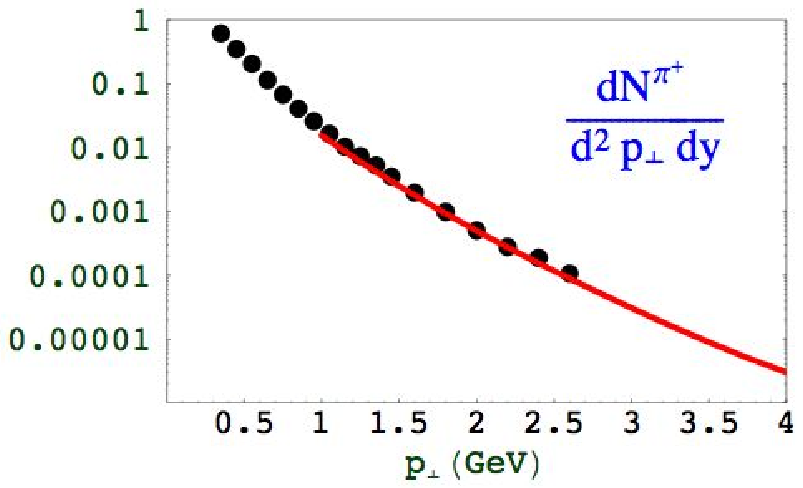}
\end{center}
\end{minipage}
\hfill
\begin{minipage}{0.32\textwidth}
\begin{center}
\includegraphics[width=\textwidth]{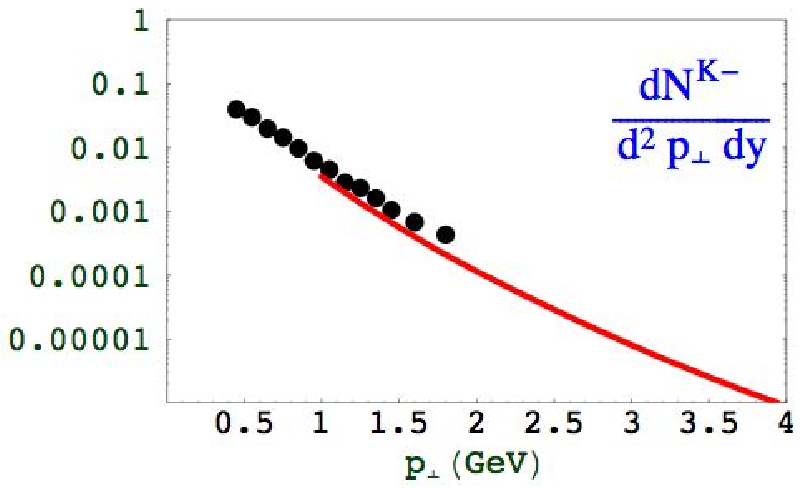}
\end{center}
\end{minipage}
\hfill
\begin{minipage}{0.32\textwidth}
\begin{center}
\includegraphics[width=\textwidth]{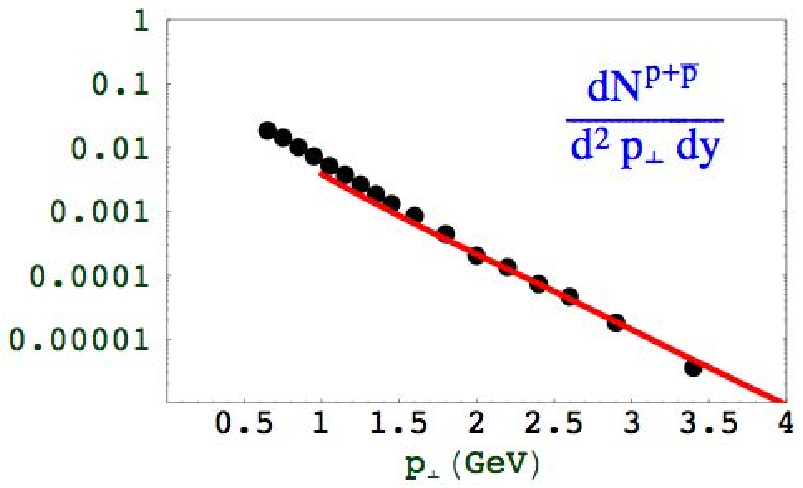}
\end{center}
\end{minipage}
\caption{\footnotesize The comparison of the spectra of produced $\pi^+$, $K^-$  and $p+\bar p$ in proton-proton collisions at $\sqrt{s}=200$ GeV with STAR experimental data from \protect\cite{Adler:2006xd}}
\label{fig:specpp}
\end{figure}

For the calculation of $R_{AA}$, we model the soft spectrum in the $p+p$ collision by the same Eqs. (\ref{eq:messpec})-(\ref{eq:k4q}) fitted to proton production data from Ref. \cite{Adler:2006xd}. This leads to values reported in the Table. The indicated fugacities are consistent with indications from the statistical model of hadron abundances; for strange particle production we have multiplied by an additional factor $\gamma_S=0.3$. The volume rescaling factor $\sim$ 9 found does not distinguish between a decrease in $\tau$ and/or $A_\perp$. Just to put some numbers, taking instead $R_p=1.2$ fm a value of $\tau\sim 1.6$ fm is obtained. Finally, the same flow velocity $v_\perp$ is assumed for $p+p$ and $A+A$ collisions at a given energy. These details do not affect much our results since within this setup the spectra are, anyway, dominated by the hard component in the $p_t$-range under study. 
We show in Fig. \ref{fig:specpp} a comparison with experimental data.
The agreement for both pion and proton production is very good, while 
it turns out to be worse between our $K^-$ and the data for proton-proton. This can be traced back to the behavior of the fragmentation function in the hard part of the spectrum. We did not attempt to improve on this, since we prefer to maintain all the parameters fixed in \cite{Friesvb}.

 \paragraph{Elliptic flow.} It is customary to expand the azimuthal dependence in the function $f_a$, Eq.  (\ref{quarkdist}) according to:
 \beq
 f_a=1+2v_2(p_\perp)\cos(2\phi_a)+\ldots
 \label{azaym}
 \eeq
 We see at once from Eq.~(\ref{hdistr}) that the hadron azimuthal distribution is:
 \beq
  f_H=1+2 V_2(P_\perp)\cos(2\Phi)+\ldots
  \eeq
  with:
  \beq
 V_2(P_\perp)\simeq 2v_2(\frac{P_\perp}{2})
 \eeq
for small $v_2$. For a hadron made of N constituents we recover the well-known scaling 
law \cite{ellflowscal}:
 \beq
 \frac{1}{N} V_2(P_\perp)= v_2(\frac{P_\perp}{N}).
\label{eq:efscal}
 \eeq

The observation of elliptic flow for the $f_0$ and the determination of the
value of $N$ by which the scaling rule (\ref{eq:efscal}) is obeyed would provide a very reliable determination of its quark structure, albeit a very difficult one, given the smallness of the expected value of $(V_2)_{f_0}$, see ref~\cite{Nonakaew} for a similar proposal for the hypothetical pentaquark.

 \paragraph{Parton fragmentation.} For $p_\perp$ large enough, particle production is dominated by hard processes computed by: 
\begin{equation}
  \label{eq:frac2}
  E \frac{d N_h}{d^3 P} = \sum_a \int\limits_0^1 \frac{d z}{z^2} 
  D_{a\to h}(z, Q^2) E_a \frac{d N^{\rm pert}_a}{d^3 P_a},
\end{equation}
where $D_{a\to h}(z,Q^2)$ is the fragmentation function of a parton $a$ into a hadron $h$, evaluated at $Q^2=P_\perp^2$. The corresponding perturbative spectrum of parton $a$: 
\begin{equation}
  \frac{d N^{\rm pert}_a}{d^2 p_{a,\perp} dy} \Big|_{y=0} = 
  K \frac{C}{(1+p_{a,\perp}/B)^\beta}
  \label{eq:pqcdpart}
\end{equation}
is taken from LO pQCD calculations for central collisions from \cite{ref2}, where the parameters $C$, $B$, and $\beta$ can be found. A constant $K$ factor of 1.5 is included \cite{Friesvb}.

The suppression of particles with $p_\perp\gtrsim$ 6 GeV is described by the energy loss of their parent partons due to radiative medium-induced processes 
\cite{reviewart1}. 
We adopt the simplified description in \cite{Friesvb}, whereby the energy loss is implemented by a shift in the perturbative spectrum by the amount: 
\beq
\Delta p_\perp(b,p_\perp) = \epsilon(b) \,\sqrt{p_\perp} \,\frac{ \langle L \rangle}{R_A}
\eeq
where:
\bea
&&\epsilon(b)=\epsilon_0\frac{1-\exp(\frac{b}{R_A}-2)}{1-\exp(-2)}\;,  \nonumber \\
&&\langle L\rangle =\frac{ \sqrt{R_A^2-\frac{b^2}{4}}+R_A-\frac{b}{2}}{2}\;.     
\eea
This gives a common suppression for all hadron species, in agreement with experimental data at large $p_\perp$ and with the results from \cite{Eskola:2004cr}. The corresponding value of $\epsilon_0$ for the LHC is also taken 
to satisfy the results from \cite{Eskola:2004cr} with a multiplicity enhancement from RHIC to LHC as given from \cite{Armesto:2004ud}. Notice that in the full jet quenching calculation, geometric effects -- surface dominated emission -- are very important \cite{Eskola:2004cr}. Here we just hide all these effects into the single parameter, $\epsilon_0$, which effectively describes the calculated suppression. The surface emission makes the extrapolation from RHIC to LHC rather insensitive to the actual value of the jet quenching parameter, $\hat q$, in the approach of \cite{Eskola:2004cr}. For this reason, we do not expect our main results to change due to this uncertainty.

For the fragmentation functions $ D_{a\to h}(z, Q^2)$ we use KKP \cite{kkp} for all the particles except $\Lambda+\bar\Lambda$ which is taken from DSV \cite{dsv}. These parameterizations are obtained from fits to experimental data and do not exist for the $\Xi$ or the $f_0$ as a four-quark meson, which will be discussed in the next Section. 

To normalize peripheral or proton-proton collisions to the central ones, we shall need the number of binary nucleon-nucleon collisions in Au+Au and Pb+Pb. 
The number of collisions for the first case is taken from \cite{Friesvb}. For the LHC, we have rescaled $N_{coll}$ with the inelastic cross section \cite{Eskola:2005ue}.

The adopted values for $\epsilon_0$ and $N_{coll}$ are given in Table~\ref{table1}. 

In summary, the total spectrum is just the sum of the soft part described by Eqs. (\ref{eq:messpec})-(\ref{eq:k4q}) and the hard part given by Eqs. (\ref{eq:frac2})-(\ref{eq:pqcdpart}). The relative importance between both contributions depend on the quark structure of the hadron and the fragmentation functions. In this sense, the recombination spectrum for a 4-quark state is broader than that for a normal meson. We will see in the next section that the fragmentation is also more suppressed in the first case than in the second. Both effects make the nuclear modification factors $R_{AA}$ and $R_{CP}$ to be enhanced at intermediate $p_t$ for particles with larger number of constituent quarks.

\section{Fragmentation functions}
\label{fragmentation}

The main uncertainty in the model is the construction of a fragmentation function for $f_0$ as a 4-quark state.  

The question of how a hard process fragments into the final hadrons is not fully solved theoretically. In the approximation of independent fragmentation (\ref{eq:frac2}), fits to experimental data by means of GLAP evolution equations have been provided. In these fits, experimental data from $e^+e^-\to$ hadrons, in particular from $Z^0$ hadronic decays are used. One additional difficulty comes from the kinematic constraints which, in these experiments, reduce the sensitivity to the fragmentation functions at large values of $z$. On the other hand, for hadronic collisions the relevant values are in fact $z\sim$ 0.7 
\cite{Eskola:2002kv}, due to the steeply falling perturbative spectrum (\ref{eq:pqcdpart}). 
At present, the description of experimental data in $p+p$ collisions at RHIC is rather good for mesons and reasonable for baryons. 

We do not attempt a full study of different fragmentation functions and their evolution. Instead, we make use of the known fragmentation functions for $\pi$, $K$, $p$ \cite{kkp} and $\Lambda$ \cite{dsv} and adopt simple rules to extrapolate to the yet unknown case of $f_0$ as a 4-quark state, comparing when possible with the scanty available data. We consider the case of $\Xi^-$+${\bar \Xi}^+$ as well, for which preliminary data are available from RHIC.

The LEP Collaborations have measured, see e.g. \cite{Ackerstaff:1998ue}, the Z$^0 \to f_0$ fragmentation function:
\beq
\frac{1}{\Gamma_{hadr}}\frac{d\Gamma_{f_0}}{dz}=\sum_{q=u,d,s}\frac{2\Gamma_q}{3\Gamma_s+2\Gamma_u}D_{q}^{f_0}(z,M_Z^2)\;.
\eeq
The fragmentation function for q$\to f_0\; (q=u, d)$ at $Q^2\sim (5\;{\rm GeV})^2$ has been measured in $\nu_\mu$-nucleus scattering by NOMAD \cite{NOMADfrag}.

We make the assumption that the fragmentation functions at large $z$ are dominated by the quark content of the particle. 
In addition, we introduce a suppression factor for each additional strange quark with respect to $\Lambda$, which we estimate
to be $\simeq 0.5$ from $K/\pi$ fragmentation ratio of KKP \cite{kkp}.
This leads to: 
\begin{eqnarray}
&&D_{d,s}^{\Xi^-+\bar \Xi^+}(z,Q^2) \sim 0.5\times D_{d,s}^{\Lambda+\bar \Lambda}(z,Q^2) \nonumber\\
&&D_u^{\Xi^-+\bar \Xi^+}(z,Q^2) \ll  D_u^{\Lambda+\bar \Lambda}(z,Q^2)\;, \nonumber
\label{eq:fragxi}
\end{eqnarray}
since $\Xi^-$ contains $d$ and $s$ but no $u$ valence quarks.

In the case of the $f_0$ as a 4-quark state, we impose an additional suppression factor $(1-z)^{1.5}$. Intuitively, this term corresponds to the suppression for producing and additional diquark with $z\to 1$. This suppression factor is present in the numerical comparison between $p$ and $\pi$ in the KKP fragmentation functions \cite{kkp}, almost independent of the $Q^2$ value, and has also been proposed in terms of Regge theory \cite{Kaidalov:1986wq}. For the gluon$\to f_0$ fragmentation function we take a larger suppression factor, $0.1(1-z)^5$, although this gluon contribution is almost irrelevant.  In conclusion we take ($q=u, d, s$):
\begin{eqnarray}
&&D_q^{f_0(4q)}(z,Q^2) \sim  0.5(1-z)^{1.5}~\frac{D_q^{\Lambda+\bar \Lambda}(z,Q^2)}{2}\nonumber\\
&&D_g^{f_0(4q)}(z,Q^2) \sim 0.1 (1-z)^{5}~\frac{D_q^{\Lambda+\bar \Lambda}(z,Q^2)}{2}
\label{frag:f0}
\end{eqnarray}
Finally, for the $f_0$ as a $s\bar s$ state, we scale $D_{q}^{K_S}$ from \cite{kkp} by:
\beq
D_{q,g}^{f_0,(s\bar s)}(z,Q^2) \sim 0.5\times D_{q,g}^{K_S}(z,Q^2)\;.
\eeq

\begin{figure}[htb]
\begin{center}
\epsfig{
height=5truecm, width=7truecm,
        figure=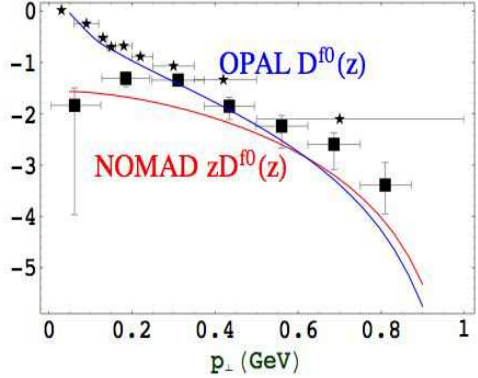}
\caption{\footnotesize 
the $Z^0\to f_0$(4q) fragmentation (\ref{eq:fragxi}) at $Q^2=10^4$~GeV$^2$ (blue curve) compared with OPAL data \cite{Ackerstaff:1998ue} (stars). In the same figure, we show $zD_q^{f_0(4q)}(z)$ at $Q^2=25$~GeV$^2$ (red curve) compared to NOMAD data (boxes). The drop we have assumed for $z\to 1$ is not incompatible with present data, within errors. 
}
\label{fragf0}
\end{center}
\end{figure}

We show in Fig.~\ref{fragf0} the $Z^0\to f_0$(4q) fragmentation (\ref{eq:fragxi}) at $Q^2=10^4$~GeV$^2$ (blue curve) compared with OPAL data \cite{Ackerstaff:1998ue} (stars). In the same figure, we show $zD_q^{f_0(4q)}(z)$ at $Q^2=25$~GeV$^2$ (red curve) compared to NOMAD data (boxes). Within errors, the drop we have assumed for $z\to 1$ is not incompatible with present data. 

Summarizing, the results -- to be presented in the next section -- are sensitive to the behavior of the fragmentation functions at large values of $z\sim 0.7$ \cite{Eskola:2002kv}, which control, in particular, the relative normalization of the hard and soft spectra for each particle species. Our strategy was to construct the $f_0$ fragmentation functions from the known functions by applying the rules given above. We have checked that the results are still compatible with the small amount of available experimental data. We have a check on the quality of this procedure in the case of the $\Xi$, where no fragmentation functions are available and the description of the data, applying the same rules as for the $f_0$, is good -- see Fig. \ref{RAA:RHIC}.

\section{Results}

\begin{figure}[htb]
\begin{minipage}[t]{75mm}
\epsfig{
height=5.3truecm, width=8truecm,
       figure=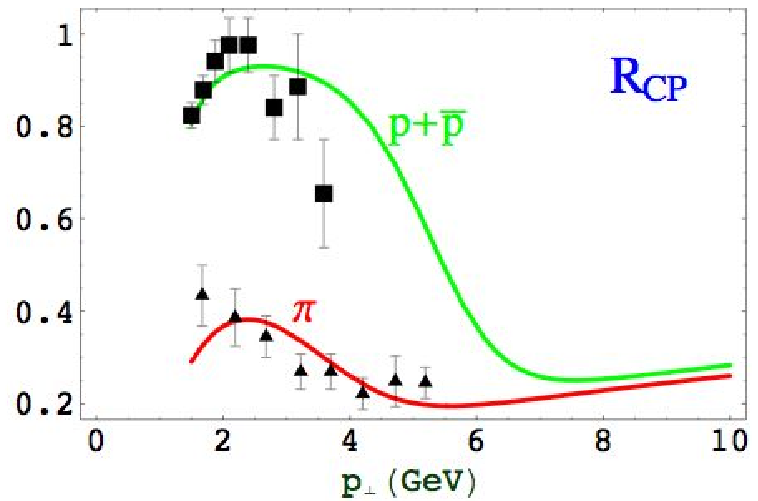}
\caption{\footnotesize R$_{CP}$, Eq.~(\ref{RCP}), for $\pi^0$ and $p+\bar p$ in Au+Au at RHIC. Peripheral collisions at $b=12$~fm. Data from~\cite{phenpi}.}
\label{RCPpi:RHIC}
\end{minipage}
\hspace{1.5truecm}
\begin{minipage}[t]{75mm}
\epsfig{
height=5.3truecm, width=8truecm,
        figure=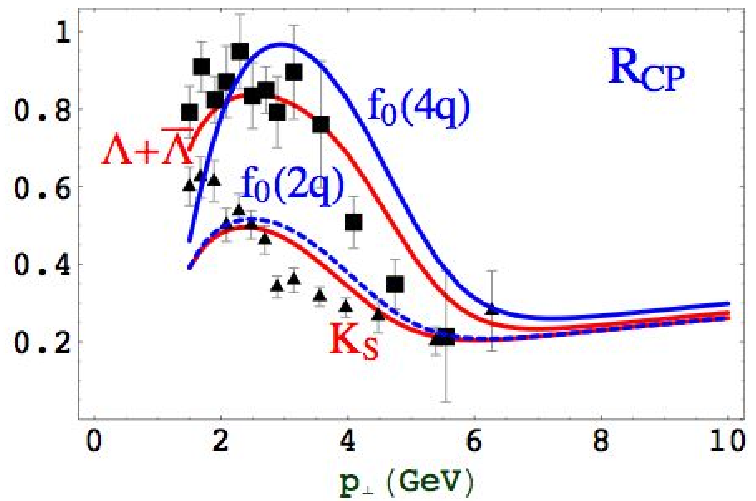}
\caption{\footnotesize {Red curves: R$_{CP}$ for $K_S$ and $\Lambda + \bar \Lambda$, Au+Au at RHIC. Peripheral collisions at $b=12$~fm. Blue curves: predictions for $f_0=s\bar s$ (dashed) or $f_0=[qs][\bar q\bar s]$ (full). Data from~\cite{phenks}.}}
\label{RCPks:RHIC}
\end{minipage}
\end{figure}

\begin{figure}[htb]
\begin{minipage}{0.49\textwidth}
\begin{center}
\includegraphics[width=\textwidth]{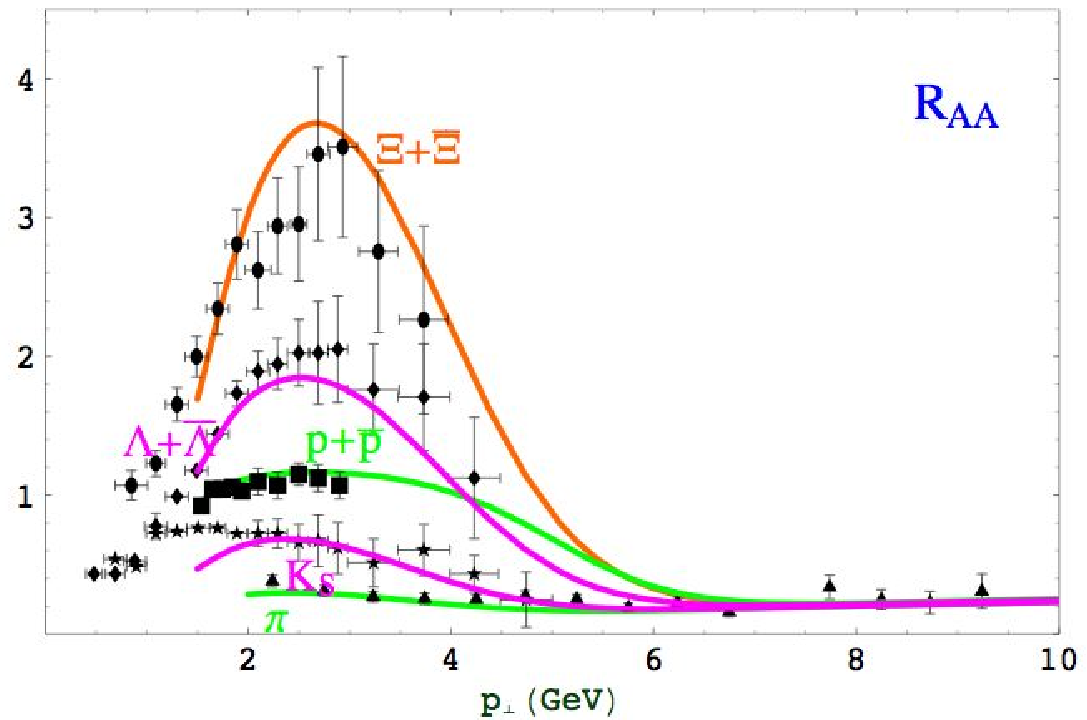}
\end{center}
\end{minipage}
\hfill
\begin{minipage}{0.49\textwidth}
\begin{center}
\includegraphics[width=\textwidth]{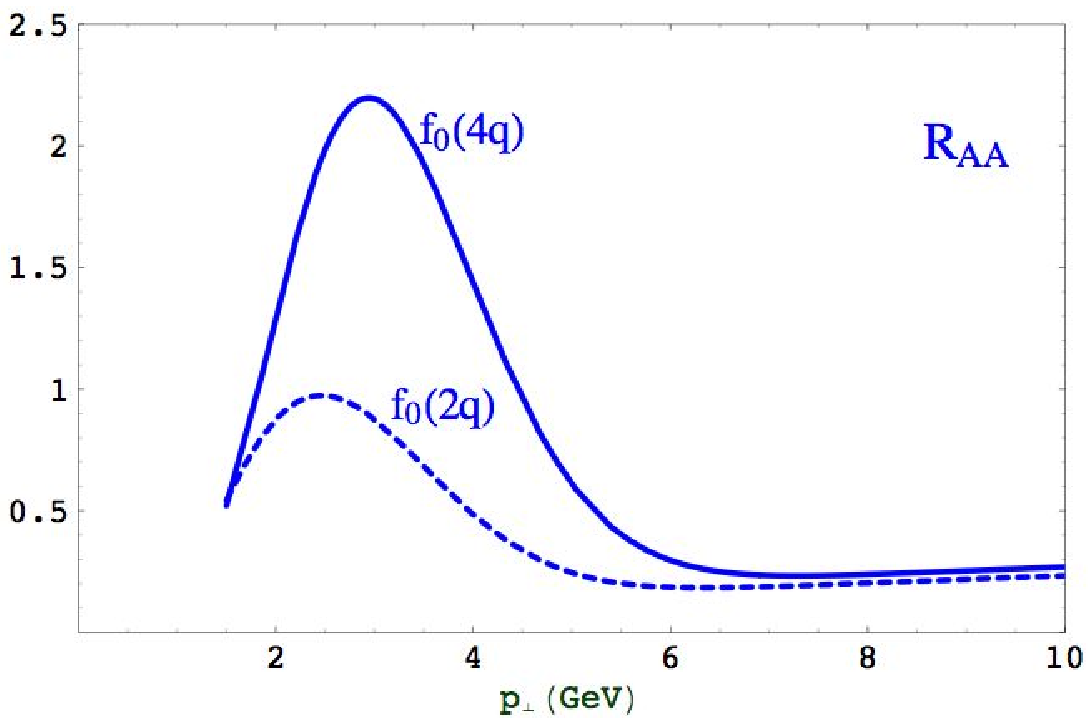}
\end{center}
\end{minipage}
\caption{\footnotesize 
Starting from below: $R_{AA}$, Eq.~(\ref{RAA}), for $\pi$, $K_S$, $f_0(980)$ as a $s\bar s$ state, $p+\bar p$, $\Lambda+\bar \Lambda$, $f_0(980)$ as a 4-quark state and $\Xi^-+{\bar \Xi}^+$. Data from Ref. \cite{Salur:2005nj}.
}
\label{RAA:RHIC}
\end{figure}

We present in Figs.~\ref{RCPpi:RHIC},\ref{RCPks:RHIC} and Fig.~\ref{RAA:RHIC} our results for the nuclear modification factors $R_{CP}$ and $R_{AA}$ at RHIC. The nuclear modification factors are defined as: 
\begin{eqnarray}
\label{RCP} 
&&R_{CP} = \frac{N_{\rm coll}(b)\, d^2N_{\rm Au+Au}(b=0)/dP_\perp^2}
{N_{\rm coll}(b=0)\, d^2N_{\rm Au+Au}(b)/dP_\perp^2},\\
&&R_{AA} = \frac{d^2N_{\rm Au+Au}(b=0)/dP_\perp^2}{N_{\rm coll}(b=0)\, d^2N_{{p+p}}/dP_\perp^2}
\label{RAA}
\end{eqnarray}
We take $b=12$ for peripheral collisions at RHIC. 

For $\pi$, protons, K and $\Lambda$ we reproduce the results of Ref. \cite{Friesvb}. The large difference between $R_{CP}$ and $R_{AA}$ for the strange baryons is, in this approach, identified to be due to the strangeness suppression factor in $p+p$, as it was also found for the total yields at SPS energies \cite{Capella:1999cz, Vance:1999pr}, and in agreement with \cite{Pop:2004dq}. 

Figs.~\ref{RCPks:RHIC} and \ref{RAA:RHIC} illustrate clearly our main result, namely that there is a large difference between the suppression for the $f_0(980)$ when assumed to be a four-quark state or a quark-antiquark meson. For both $R_{CP}$ and $R_{AA}$, the nuclear modification ratio for the 4-quark state is in the range of the strange baryon ratio, while it is closer to the K$_S$ signal for a $s\bar s$ state. The difference is particularly large in the experimentally accessible range of $p_\perp\lesssim$ 6 GeV. 

Taking our calculations at face value, the effect is larger in the case of $R_{AA}$, but the uncertainties are also larger, due to the implementation of the unknown fragmentation functions for these mesons, which in the case of $R_{CP}$ cancel to some extent. Nonetheless, even if the effect could be overemphasized in the present formalism, we expect $R_{AA}$ for the four-quark state to be in the range of the corresponding values for $\Lambda+\bar \Lambda$ and $\Xi^-+\bar \Xi^+$. This gives still a large leeway to distinguish the valence quark structure of the $f_0(980)$.

\begin{figure}[htb]
\begin{minipage}[t]{75mm}

\epsfig{
height=5.0truecm, width=7.7truecm,
       figure=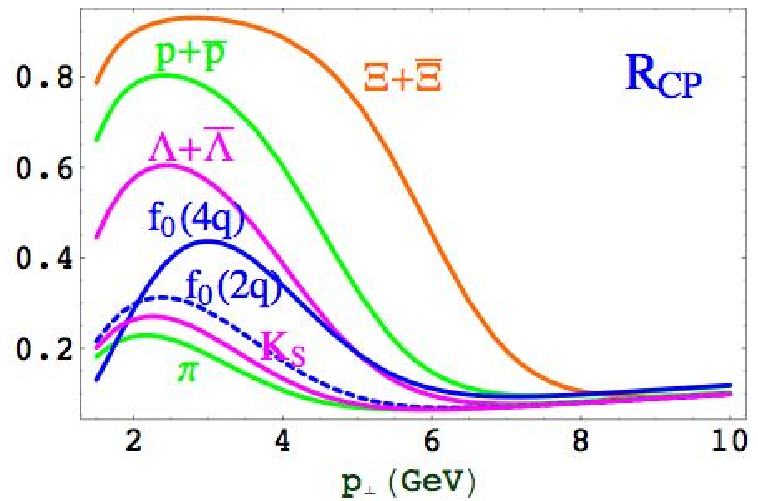}
\caption{\label{fig RCPLHC} \footnotesize 
$R_{CP}$ for LHC using the same notation as in Fig.~\ref{RAA:RHIC}.
}
\label{RCP:LHC}
\end{minipage}
\hspace{1.5truecm}
\begin{minipage}[t]{75mm}
\epsfig{
height=5.3truecm, width=8truecm,
        figure=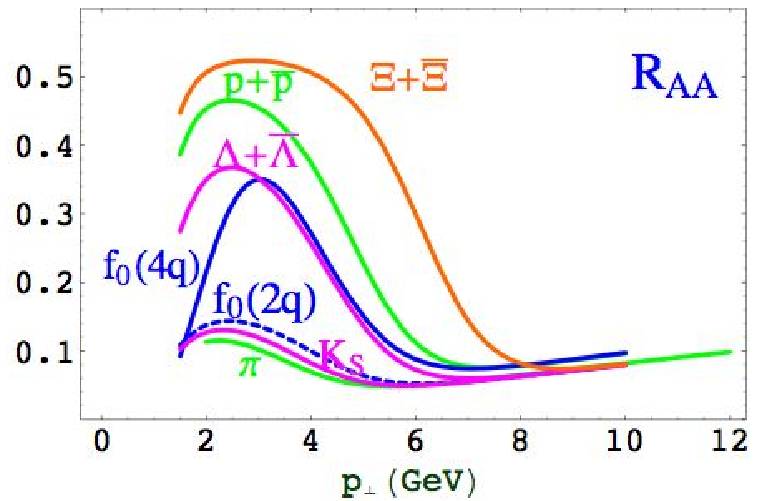}
\caption{\label{fig TH} \footnotesize 
$R_{AA}$ for LHC using the same notation as in Fig.~\ref{RAA:RHIC}.
}
\label{RAA:LHC}
\end{minipage}
\end{figure}

To extend the formalism to the case of the LHC we extrapolate the parameters as discussed in Section~\ref{model}, both for the hard part of the spectrum, given by fragmentation, and for the soft part. For peripheral collisions we take $b=14$~fm. The corresponding predictions with this set of extrapolating parameters are presented in Figs.~\ref{RCP:LHC} and \ref{RAA:LHC}. 

The R$_{CP}$ and R$_{AA}$ signals are considerably reduced, presumably due to the large increase of gluon and sea-quark densities in the proton, which entails an increase in the fragmentation component. This conclusion, however, is strongly sensitive to the assumed value of the recombination volume, $\tau A_\perp$, of the fugacities at LHC as well as to the values of $v_\perp$ \cite{Fries:2003fr}. Cross section data are needed for a more reliable assessment of the parameters of the model, as is the case at RHIC. Nonetheless, the correlation of the R$_{CP/AA}$ ratios of a 4-quark $f_0$ with the baryon ratios is still borne out by our calculation.

\section{Discussion}

We have presented experimentally observable quantities, which in the region of intermediate transverse momentum in heavy ion collisions are sensitive to the constituent quark content of the resonances. For this purpose, we use a simple but successful model based on a hadronization by recombination of the quarks in the fireball produced after a heavy ion collision \cite{Friesvb}. We extend this model to the multi-strange and multi-quark hadrons, finding a good description of both $R_{AA}$ and $R_{CP}$ with the usual assumptions on strangeness suppression. We extrapolate the model to the LHC by fitting to the available results in more refined calculations of both the soft \cite{Eskola:2005ue} and the hard contributions to the spectrum. Uncertainties however remain in the values of the parameters at the LHC, which will be possible to eliminate when cross section data will become available.

The $f_0(980)$ resonance is, at present, the best possible case for these type of studies as (i) the signals are large enough for an unambiguous result to be obtained and (ii) the resonance is abundantly  enough produced \cite{Adams:2003cc} to afford a measurement of the nuclear modification ratios.

\begin{figure}[htb]
\begin{center}
\epsfig{
height=4.7truecm, width=7.9truecm,
        figure=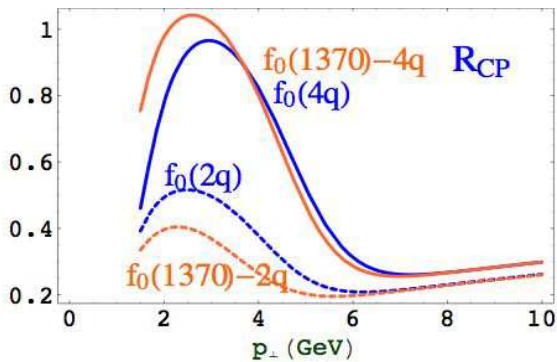}
\caption{\footnotesize $R_{CP}$ for LHC. We compare the two hypotheses,
$2q$-dashed and $4q$-solid, for the $f_0(980)$ and $f_0(1370)$.  
The four quark interpretation of  $f_0(1370)$ is $[qq][\bar q\bar q]$, $q$ being a light
non-strange quark~\cite{noilast}
}
\label{f1370}
\end{center}
\end{figure}

Other possibilities concern the $f_0(1370)$ and $f_0(1700)$ mesons, for which reasons to suspect a 4-meson interpretation have been recently advanced \cite{noilast}, alternative to the conventional interpretation as P-wave $q\bar q$ states \cite{frankclose1}. We do not have any information on fragmentation functions into these mesons analogous to those provided by OPAL and NOMAD, see Fig.~\ref{fragf0}. To give an idea anyway, we simply scale with the strange quark content, taking for $f_0(1700)$ the same as in (\ref{frag:f0}) and multiplying by $1/(0.5)^2$ for $f_0(1370)$~\cite{noilast}. We report in Fig.~\ref{f1370} the resulting values of $R_{CP}$ at RHIC of the $f_0(1370)$ according to the two assumptions on composition, compared to the analogous curves for $f_0(980)$. The predictions for $f_0(1700)$ essentially coincide with those for $f_0(980)$.

In the same mass region, the $f_0(1500)$ scalar meson has been long suspected \cite{frankclose1} to be the lightest glueball predicted by QCD lattice calculations (see e.g. \cite{glueball}). We may advance two possibilities. One is that recombination of {\it constituent gluons} exists. Since the lightest glueball should be made by two constituent gluons, the $f_0(1500)$ would then follow a nuclear modification pattern close to that of conventional mesons. Alternatively, glue recombination could be suppressed and the $f_0(1500)$ signal arise from its mixing with the $f_0(1370)$ and $f_0(1700)$ mesons only. The $f_0(1500)$ would exhibit, in this case, an anomalous suppression in the coalescence region.

Experimental investigations at RHIC and LHC could shed light on the above issues and produce a decisive advance in our understanding of meson spectroscopy.


\section*{Acknowledgements}
We wish to thank N.~Bianchi, M.~Cacciari and T.~Sj\"{o}strand for informative conversations. CAS is supported by the 6th Framework Programme of the European Community under the 
Marie Curie contract MEIF-CT-2005-024624.

\end{document}